\documentclass[sn-mathphys,Numbered]{sn-jnl}

\usepackage{graphicx}%
\usepackage{multirow}%
\usepackage{amsmath,amssymb,amsfonts}%
\usepackage{amsthm}%
\usepackage{mathrsfs}%
\usepackage[title]{appendix}%
\usepackage{xcolor}%
\usepackage{textcomp}%
\usepackage{manyfoot}%
\usepackage{booktabs}%
\usepackage{algorithm}%
\usepackage{algorithmicx}%
\usepackage{algpseudocode}%
\usepackage{listings}%
\usepackage{tabularx}%

\theoremstyle{thmstyleone}%

\theoremstyle{thmstyletwo}%

\theoremstyle{thmstylethree}%

\begin{document}

\title[Article Title]{Aalap: AI Assistant for Legal \& Paralegal Functions in India}

\author*[2]{\fnm{Aman} \sur{Tiwari}}\email{aman.tiwari@thoughtworks.com}

\author[2]{\fnm{Prathamesh} \sur{Kalamkar}}\email{prathamk@thoughtworks.com}
\equalcont{These authors contributed equally to this work.}

\author[1]{\fnm{Atreyo} \sur{Banerjee}}\email{atreyo@agami.in}
\equalcont{These authors contributed equally to this work.}

\author[1]{\fnm{Saurabh} \sur{Karn}}\email{karnsaurabhkumar@gmail.com}

\author[1]{\fnm{Varun} \sur{Hemachandran}}\email{varun@agami.in}

\author[1]{\fnm{Smita} \sur{Gupta}}\email{smita@agami.in}

\affil*[1]{ \orgname{Agami}}
\affil*[2]{ \orgname{Thoughtworks Technologies India Pvt Ltd}}

\abstract{
Using proprietary Large Language Models on legal tasks poses challenges due to data privacy issues, domain data heterogeneity, domain knowledge sophistication, and domain objectives' uniqueness. We created Aalalp, a fine-tuned Mistral 7B model on instructions data related to specific Indian legal tasks. The performance of Aalap is better than gpt-3.5-turbo in 31\% of our test data and obtains an equivalent score in 34\% of the test data as evaluated by GPT4. Training Aalap mainly focuses on teaching legal reasoning rather than legal recall.
Aalap is definitely helpful for the day-to-day activities of lawyers, judges, or anyone working in legal systems.
}

\maketitle

\section{Introduction}
Large Language Models (LLMs) have triggered the creation of innovative applications in the legal field. 
While many LLM-powered applications in the legal domain use cloud-hosted LLMs like OpenAI, Bard, etc., there are some challenges.
Data privacy issues and regulatory requirements often require that data can't leave an organization's boundary. This may require an LLM to be hosted on-premise within the organization. Hosting larger models with tens of billions of parameters can be cost-prohibitive. Hence, there is a need to distill the knowledge of larger models into smaller models that work well on specific legal tasks. While state-of-the-art models are very good at bar exams and other objective legal assessments \cite{Katz_Bommarito_Gao_Arredondo_2023}, there is still room for improvement.  \cite{guha2023legalbench}, \cite{Choi_Hickman_Monahan_Schwarcz_2023} and \cite{chalkidis2023chatgpt} show areas where the gaps of state-of-the-art LLMs for legal tasks.


Fine-tuning a general-purpose LLM on specific legal tasks can alleviate some of these concerns. Hence, we created \textbf{Aalap}, which stands for AI Assistant for Legal and Paralegal functions in India. The following are the main contributions of this paper.
\begin{itemize}
    \item \textbf{Aalap instructions dataset}\footnote{\url{https://huggingface.co/datasets/opennyaiorg/aalap_instruction_dataset}}: 22272 legal instructions and synthetically generated responses with a maximum input token length of 16k
    \item \textbf{Aalap model} \footnote{\url{https://huggingface.co/opennyaiorg/Aalap-Mistral-7B-v0.1-bf16}}: 7B Mistral fine-tuned on the Aalap dataset with a maximum context length of 32k which can perform well for specific legal tasks in Indian context.
    \item \textbf{AIBE data} \footnote{\url{https://huggingface.co/datasets/opennyaiorg/aibe_dataset}}: AIBE dataset containing 1158 multiple-choice questions and answers from past 12 AIBE exams.
\end{itemize}

Aalap model weights and the dataset are available on huggingface.

\section{Related Work}
Domain adaption of LLMs can be done in three broad ways \cite{ling2023domain}. The first one is pretraining from scratch on domain-specific raw text datasets. Some examples of this approach are BloombergGPT \cite{wu2023bloomberggpt} and Med-PaLM2 \cite{singhal2023expertlevel}. While this approach seems promising, it needs a lot of computing power and data as guided by Chinchilla scaling laws \cite{hoffmann2022training}. The second approach is fine-tuning an LLM using domain-specific supervised datasets. This approach needs less computing power and data than pre-training from scratch. Such smaller fine-tuned models can sometimes match the performance of much larger models on specific tasks \cite{cheng2023adapting}.
approach. The third approach is using Retriever Augmented Generation \cite{lewis2021retrievalaugmented} \cite{cui2023chatlaw}, which augments the prompt of an LLM with retrieved domain, and no pretraining or fine-tuning is involved. This method also involves prompt engineering techniques that guide LLMs to accomplish specific tasks. 

Domain adaptation of general-purpose LLMs to law via supervised fine-tuning has been widely studied. Notable ones are LlawyerLlama \cite{huang2023lawyer}, a Chinese legal LLM trained on a large-scale legal dataset, and ChatLaw \cite{cui2023chatlaw}, a Legal Large Language Model with Integrated External Knowledge Bases. Both models rely on retrievers to extract relevant external knowledge and generate the required answers using LLM abilities. 

Prompting general-purpose LLMs using legal reasoning frameworks like IRAC (Issue, Rule, Application, Conclusion) coupled with few-shot examples is an effective way to extract more from LLMs without the need of fine-tuning \cite{yu2023exploring}. \cite{trautmann2022legal} have used legal prompt engineering for Multilingual Legal Judgement Prediction. 

Evaluation of law-specific LLMs needs benchmarks that focus on legal tasks. Since variations in legal processes by country are significant, developing a single global legal LLM evaluation benchmark is hard. Legalbench is a collaboratively constructed legal reasoning benchmark comprising 162 tasks in the USA law context \cite{guha2023legalbench}. LawBench \cite{fei2023lawbench} is a similar benchmark in the Chinese context. However, such legal benchmarks in the Indian context do not exist to the best of our knowledge.

Legal texts are usually longer; tasks like argument generation need input information from multiple documents. Hence, a legal LLM needs to handle longer input texts. Recent advances in training and inference of LLMs allow for handling longer texts at lesser compute. The most significant advances are Grouped Query Attention \cite{ainslie2023gqa}, and Sliding Window Attention \cite{child2019generating}. Mistral 7B \cite{jiang2023mistral} is an open-source LLM that has leveraged such advances and performs well on a wide range of benchmarks. Instruction fine-tuning methods like LoRA \cite{hu2021lora} and QLoRA \cite{dettmers2023qlora} allow for parameter-efficient fine-tuning of LLMs.

\section{Aalap Dataset}
The main objective behind creating Aalap was to train a relatively smaller LLM specializing in specific legal tasks focusing on legal reasoning. Hence, prioritizing which legal tasks to focus on was an important step.
We discussed with Legal practitioners which legal tasks they want to be automated. Then, we selected the most common legal tasks for publicly available input datasets. The selected legal tasks are mentioned in table \ref{legal_tasks}

We followed the following principles during data creation
\begin{enumerate}

    \item \textbf{Focus on legal reasoning rather than legal recall}: Relying on LLMs to generate legal precedents and statute definitions is a bad idea. This is because LLMs are more likely to hallucinate and generate false content. Instead, precedents information and statute definitions should be retrieved from authentic sources, given as input to LLMs, and let LLMs do the legal reasoning. 
    \item \textbf{Use real-life situations}: The datasets should contain real-life situations as much as possible. Hence, we used descriptions from court judgments and police First Investigation Reports (FIR) so that the model learns the language used in such texts.
    \item \textbf{Explain answers}: ORCA \cite{mukherjee2023orca} style explanations of answers help teach models the reasons behind coming up with answers to multiple-choice questions. We created GPT4 to create explanations or reuse relevant legal datasets that provided such explanations. 
    \item{\textbf{Use synthetically generated responses if needed}}: Many times, the actual documents, like lawyers' pleadings, witness statements, medical reports, etc., are not available publicly. So, synthetically generated responses from OpenAI models were used wherever needed.
\end{enumerate}

\subsection{Common Data Sources}
The following are the commonly used data sources for preparing the input instructions for Aalap dataset.
\begin{itemize}
    \item \textbf{Judgments Facts}: Case facts are typically written in a court judgment. Such facts from judgments texts were extracted using Opennyai's rhetorical roles model \cite{kalamkarcorpus} on randomly chosen Indian Supreme Court \& high court judgments. These judgment facts were inputs for tasks like argument generation, issues generation, and event timeline creation.
    \item \textbf{Judgments Statutes}: Relevant Statutes of a judgment are the ones that were extracted from corresponding judgment text using Opennyai's Named Entity Recognition model \cite{kalamkar2022named}.
    \item \textbf{FIRs}:
    Publicly available police First Investigation Reports (FIRs) from the states of Maharashtra and Delhi were used to collect descriptions of real-life events. These FIRs were used to create event timeline creation tasks.
\end{itemize}

\subsection{Data Creation}
The datasets for prioritized legal tasks were created as described in table \ref{legal_tasks}.

\begin{table}[ht!]
\centering
\hskip-10.0cm
\begin{tabular}{| m{5em} | m{5.25cm}| m{5.25cm} |}
    \hline

        \textbf{Task} & \textbf{Task Description} & \textbf{Data Creation Methodology} \\ \hline
      
        Issues Generation & Create legal issues for a court based on the case facts. Legal issues are the key points on which the verdict needs to be delivered. & Judgment facts were then sent to gpt-3.5-turbo to generate legal issues. \\ \hline
        
        Argument Generation & Based on the facts of a case, legal issues, and applicable statute names, generate arguments for a given party.  & Arguments for the petitioners were created using gpt-3.5-turbo, using judgment facts, generated issues, and relevant statutes as inputs. Counterarguments for defendants were created using petitioners' generated arguments. \\ \hline
        
        Event Timeline & Extract important events and their dates from the input text descriptions and output a chronologically sorted event list with dates and a brief description. & FIRs and judgment facts were used as input text descriptions. These were sent to gpt-3.5-turbo to create event timelines. \\ \hline
        
        Combine Event Timelines & For extraction of event timelines from very long texts, it is often split into chunks, and the event timeline for each chunk is created independently, which is merged later. & Individually created timelines from the same judgment or FIR were merged using gpt-3.5-turbo. \\ \hline
        
        Legalbench & Training data for the legalbench \cite{guha2023legalbench} was filtered to keep only the legal reasoning tasks. &  ORCA-style explanations of these MCQs were created using GPT4 in a zero-shot setting. \\ \hline
        
        Statute Ingredients & Break the input statute definition into the ingredients needed to apply the statute. & Definitions of the most popular sections of Indian central acts were used to generate the statute ingredients using gpt-3.5-turbo.\\ \hline
        
        Summary Generation & Create a summary in judgment headnotes format using the input court judgment text  & Indian Supreme Court judgments from 1950 to 1994 are published with headnotes, which are summaries of those judgments. \\ \hline
        
        Legal Open ORCA & OpenORCA dataset \cite{OpenOrca} is an ORCA-style explanation of the Natural Instructions dataset. & The Natural Instruction dataset (NI) is filtered for law and matched against the 1M GPT4 openORCA dataset using tf-idf matching to get legal OpenORCA data.\\ \hline
        
        Contract Clause Generation & Generation of new contract clauses and modification of existing contract clauses. &  Existing Data\footnotemark[1]  \\ \hline
        
        Legal NIv2 MCQ & Natural Instructions v2 data \cite{naturalinstructions} was filtered for law-related questions. & A random sample of NI dataset MCQs filtered for law. \\ \hline
        
        Constitution General Knowledge & Q\&A about the Indian Constitution. &  Existing data\footnotemark[2]   \\ \hline
        
        Incomplete Instructions & The given information is incomplete for completing the task. In such cases, the response is to ask for the required information. & Randomly selected instructions belonging to each of these tasks above where the information is incomplete.\\ \hline
    
    \end{tabular}
\footnotetext[1]{\url{https://huggingface.co/datasets/NebulaSense/Legal_Clause_Instructions}}
\footnotetext[2]{\url{https://huggingface.co/datasets/nisaar/Constitution_of_India}}
\caption{\label{legal_tasks}
Prioritized Legal Tasks
}
\end{table}
Summary statistics of various task categories and licenses associated with each dataset are shown in Table \ref{table: task_summary}.

\subsection{Data Limitations}
Precedent Information is missing in the argument generation. Precedents are essential elements to support generated arguments. Fetching the proper paragraphs from authentic precedent sources based on the situation is necessary to build an argument. Since we did not have access to a semantic precedents search engine that can do this, we have excluded precedents information from generated arguments.

Facts in the judgments can be worded much differently than what is available from statements, filings, and other reports that lawyers get as inputs. Since this information is not public, we had to rely on publicly available datasets.

Since the Indian judiciary operates mainly in English, all the datasets are currently in the English language.

\section{Model Training}
To create a general-purpose Large Language Model (LLM) tailored for specific legal tasks, a model architecture with a substantial context length of 32K, namely Mistral 7B \cite{jiang2023mistral}, was chosen. This intentional selection is derived from our dataset's distinctive nature and extensive length as well as Mistral 7B's superior capabilities when compared to other models, including the LLama2 family. However, the fine-tuning of such a model necessitates considerable computational resources and comprehensive training data, as per conventional approaches. In light of the constraints imposed by limited computational resources and the specificity of our legal tasks-related dataset, it becomes imperative to adopt methods prioritizing efficiency in both computational costs and data requirements. This goal is achieved by utilizing parameter-efficient tuning methods \cite{li2021prefix} \cite{hu2021lora}, which better leverage available data and minimize the necessity for extensive resource allocation. Specifically, the Low-Rank Adaption method \cite{hu2021lora} was employed to fine-tune the Mistral 7B \cite{jiang2023mistral} model for developing Aalap.

\subsection{Training Procedure}
The fine-tuning process was conducted by loading the Mistral 7B model in Bfloat-16 precision. The training infrastructure utilized 4 X A100 80GB GPUs, affording a compute size of 320GB GPU memory resources. Considering limited resources, parameters that fully leveraged the available compute were selected, resulting in generating a commendable model within a reasonable time frame of \textbf{86 hours}, which cost us \$950.

Throughout the training phase, the maximum token length was designated as 16384, aligning with the maximum length of data present in the training corpus. The LoRA rank was initialized to 8, and alpha was set to 16, with a dropout rate of 0.1. LoRA was applied to all model layers to achieve performance similar to that of a fully fine-tuned model. Integration of flash-attention 2 \cite{dao2023flashattention2} during the training process significantly enhanced training speed. A Paged Adam 32-bit optimizer with a learning rate of 5e-5 was employed, along with a "constant" learning rate scheduler. Weight decay was set to 0.01, and the warm-up ratio was established at 0.03. Deepseed stage 3 was employed in the training process, with per\_device\_batch\_size to 4 and gradient accumulation set to 2, resulting in a global batch size of 32 across 4 GPUs. The model was trained for a total of 6500 steps which come down to a total of 10 epochs.

Our runs were tracked through Weights \& Biases \cite{wandb}. The progression of model training and evaluation losses during the fine-tuning process is depicted in Figure \ref{fig:train_eval_loss}. It is observed that there is a steep decline in the initial phase of training, followed by a slow but constant decrease in losses as the training proceeds. This steep decline suggests that, despite differences between the task and the pretraining of the model, the model demonstrates the ability to adapt quickly. As the training progresses, the model appears to focus on learning the subtle nuances of the data.

Throughout the training process, checkpoints of the model are consistently extracted for human evaluation. As the training continues, model showcases a noticeable improvement in its capability of generating responses that are not only coherent but also contextually correct for the specific legal tasks under consideration. A detailed discussion of the evaluation of Aalap is provided in the following section.

\section{Model Evaluations}
The aalap model was evaluated using three methodologies.
\begin{itemize}
    \item Using GPT4 as an evaluator on Aalap test data
    \item Using Legalbench data
    \item Using All India Bar Exam (AIBE) data
\end{itemize}

\subsection{Evaluation using GPT4 as an evaluator on Aalap test data}
The performance of Aalap was compared with Mistral 7B \& gpt-3.5-turbo on Aalap test data. The evaluation approach proposed by \cite{zheng2023judging} was followed. The ‘gpt-4-1106-preview’ model was used for the evaluations to give a score to each response ranging from 0 to 10 by comparing it with the reference answer provided. The evaluation was done mainly for correctness and helpfulness.
\begin{figure*}[h!]
\begin{center}
\includegraphics[scale=0.8]{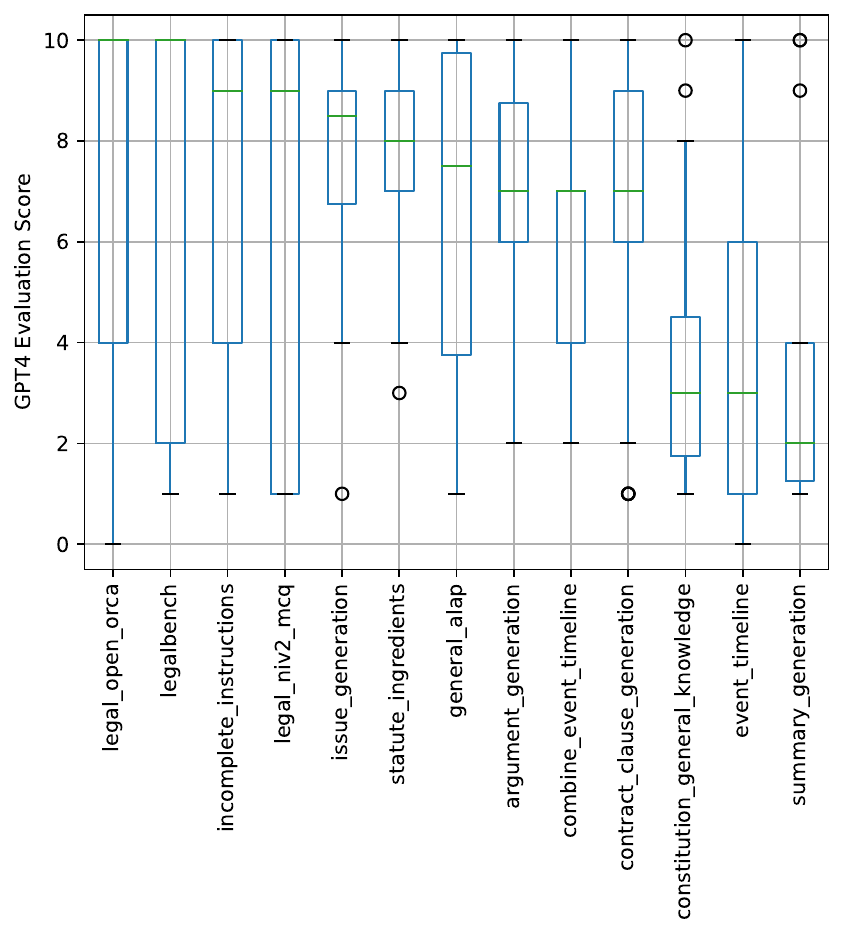} 
\caption{Boxplot of Aalap evaluation scores by GPT4 grouped by tasks}
\label{fig:rating_by_task}
\end{center}
\end{figure*}
The table \ref{aalat_results_gpt4} gives the pairwise comparison of evaluation scores.


\begin{table}[h!]
\centering
\begin{tabular}{@{}llll@{}}
\toprule
        \textbf{Model} & \textbf{Aalap is worse} & \textbf{Tie} & \textbf{Aalap is Better}\\
        \midrule
        \textbf{Mistal 7B} & 37 & 132 & 925 \\ 
        \textbf{gpt-3.5-turbo} & 385 & 373 & 336 \\ 
        \botrule
    \end{tabular}
\caption{\label{aalat_results_gpt4}
Comparison of Aalap vs. Mistral7B vs. gpt\-3.5\-turbo
}
\end{table}

We can see that Aalap is doing better than Mistral 7B in 85\% of test data. The performance of Aalap is better than gpt-3.5-turbo in 31\% of our test data and obtains an equivalent score in 34\% of the test data.

Figure \ref{fig:rating_by_task} shows the evaluation scores of Aalap calculated by GPT4 for various task categories on the test data. For verbose tasks like issues generation, argument generation, and contract clause generation, Aalap is doing a good job. We can see that Aalap is not doing well on tasks like summary generation, event timeline generation, and constitution general knowledge. 

\subsection{Evaluation using Legalbench data}
Legalbench test data \cite{guha2023legalbench} evaluates the model on 162 legal tasks, which can be categorized into Issue, Rule, Conclusion, Interpretation, and Rhetoric. Since this test data is huge, we took a random sample of 100 examples from each of the 162 tasks. The questions are mostly multiple-choice questions where the model's job is to look at a few shot examples and select the right option for the question asked. Since Aalap is trained to produce verbose output and explanations, we created longer answers for these questions and asked gpt-3.5-turbo to evaluate against ground truth. This was necessary because, many times, simple regular expressions could not evaluate the outputs correctly. For evaluating Mistral and gpt-3.5-turbo, we created the outputs with limited tokens and used regular expressions to evaluate. 

\begin{table}[h!]
\centering
\begin{tabular}{@{}llll@{}}
\toprule
        \textbf{Task Category} & \textbf{gpt-3.5-turbo} & \textbf{Mistral 7B} & \textbf{Aalap} \\
\midrule
        Issue & 51.2 & \textbf{51.8} & 45.3 \\ \hline
        Rule & \textbf{34.0} & 24.0 & 32.0 \\ \hline
        Interpretation & \textbf{61.9} & 57.0 & 53.6 \\ \hline
        Rhetoric & \textbf{62.0} & 42.0 & 33.0 \\ \hline
        Conclusion & \textbf{78.0} & 42.6 & 61.7 \\ \hline
        Overall & \textbf{61.0} & 53.2 & 51.2 \\
        \botrule
    \end{tabular}
\caption{\label{tab:legalbench_results}
The average performance of various LLMs on the sample Legalbench data
}
\end{table}
As we can see from Table \ref{tab:legalbench_results}, Aalap is doing worse than the Mistral 7B model. This is primarily due to the loss of few-shot abilities of LLMs after finetuning. This behavior is demonstrated by other studies as well \cite{wang2023twostage}. Even though the Legalbench training data is part of the training data for Aalap, the format is in the form of an explanation of an individual example explanations.

\subsection{Evaluation using All India Bar Exam}
All India Bar Exam (AIBE) questions and answers for the past 12 years were compiled \footnote{\url{https://huggingface.co/datasets/opennyaiorg/aibe_dataset}} ranging from AIBE 4 to AIBE 16. The scanned question papers with answer keys were converted to text using optical character recognition. This data was manually verified for correctness. There are 1158 multiple-choice questions from various areas of law.
We compared the performance of Aalap with Mistral 7B \& gpt-3.5-turbo on this data.  Most AIBE questions are recall-based, and a few test legal reasoning.  The minimum passing percentage for this exam is 40\%.

\begin{table}[h!]
\centering
\begin{tabular}{@{}lll@{}}
\toprule
        \textbf{gpt-3.5-turbo} & \textbf{Mistral 7B} & \textbf{Aalap}\\ 
\midrule
        \textbf{58.72}\% & 23.48\% & 25.56\% \\
\botrule
    \end{tabular}
    \caption{\label{tab:aibe_results}
Average AIBE scores for various LLMs}
\end{table}
Aalap is doing slightly better than Mistral but still falls short of the passing percentage.

\subsection{Discussion}
For the tasks that are present in the training data, we can say that Aalap is performing comparably to ‘gpt-3.5-turbo’. 
But for the AIBE exam and Legalbench data, Aalap is not doing any better than the Mistral 7B base model. The main reasons behind poor performance on AIBE data is recall-oriented nature of the questions. The main reasons for poor performance on Legalbench data are loss of few-shot abilities after fine-tuning and the dataset is in the context of US legal systems.
The performance of these benchmarks could be improved by adding such related datasets in the training process.

\section{Conclusion \& Next Steps}
We created Aalap data and model, which is a fine-tuned LLM for specific Indian legal tasks. Aalap model can do specific legal tasks well and is not a general purpose legal LLM. Given the size of the model, this can be hosted on-premise which can help with handling of sensitive data.
Hence we believe that Aalap can help to automate some mundane tasks for legal practitioners and help to improve their efficiency. 

The training data quality can be improved significantly by adding expert reviews. The entire data generation process can be more aligned with the practical workflows of lawyers by seeking their feedback. Seeking feedback about Aalap's performance in real-world scenarios will help to improve this model further. It will be useful to make this dataset \& model multi-lingual.

\section*{Acknowledgements}
This work is part of the OpenNyAI mission, which is funded by EkStep and Agami. We thank all the lawyers, including Professor Rangin Tripathi from National Law University Odisha, for contributing to the task selection. The authors thank Sankalp Bhoyar for helping build the front end for Aalap and Kushagra Bhatia for verifying and correcting AIBE dataset. We thank E2E Networks for providing GPUs for this work. 

\begin{appendices}

\section{Summary statistics of various task categories}
Summary statistics of train and test Aalap data is show in Table \ref{table: task_summary}. Average input and output tokens are calculated using tiktoken library.
\begin{table*}[h!]
\centering
\begin{tabular}{ | m{4.5cm} | m{0.8cm}| m{0.8cm} | m{1.2cm} | m{1.2cm} | m{1.5cm} |}
\hline
        \textbf{Task Category} & \textbf{train count} & \textbf{test count} & \textbf{Average input tokens} & \textbf{Average output tokens} & \textbf{License} \\ \hline
        Issue Generation & 577 & 24 & 1376 & 161 & CC0-1.0 \\ \hline
        Argument Generation & 1142 & 58 & 2381 & 943 & CC0-1.0 \\ \hline
        Event Timeline & 676 & 49 & 3476 & 342 & CC0-1.0 \\ \hline
        Combine Event Timeline & 195 & 9 & 883 & 772 & CC0-1.0 \\ \hline
        legalbench & 580 & 27 & 229 & 218 & Other \\ \hline
        Statute Ingredients & 499 & 27 & 402 & 111 & CC0-1.0 \\ \hline
        Summary Generation & 686 & 14 & 7635 & 1413 & CC0-1.0 \\ \hline
        Legal Open ORCA & 8142 & 413 & 449 & 91 & MIT \\ \hline
        Contract Clause Generation & 4325 & 232 & 76 & 179 & cc-by-nc-4.0 \\ \hline
        Legal NIv2 MCQ & 1891 & 109 & 408 & 9 & Apache 2.0 \\ \hline
        Constitution General Knowledge & 889 & 44 & 36 & 85 & Apache 2.0 \\ \hline
        Incomplete Instructions & 1464 & 82 & 97 & 81 & CC0-1.0 \\ \hline
        General Alap & 112 & 6 & 31 & 25 & CC0-1.0 \\ \hline
    \end{tabular}
\caption{\label{table: task_summary}
Summary statistics of various task categories
}
\end{table*}

\section{Aalap Training \& Eval Loss}
\begin{figure}[h]
\begin{center}
\includegraphics[width=3.5in, height=2in]{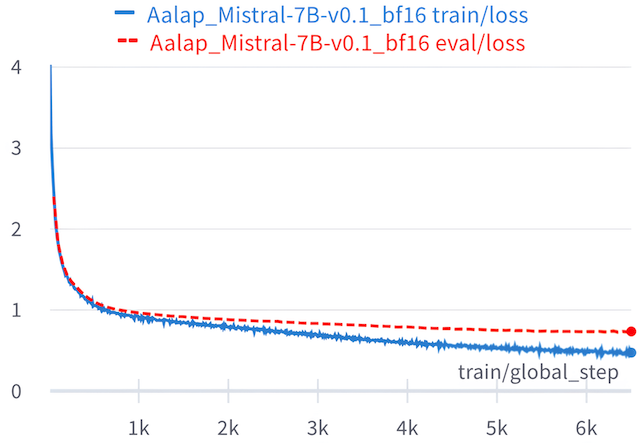} 
\caption{Comparison of Training and Evaluation Loss During Fine-tuning: The graph illustrates the progression of training and evaluation loss over the course of fine-tuning, providing insights into the model's learning dynamics and generalization performance.}
\label{fig:train_eval_loss}
\end{center}
\end{figure}

Training and evaluation loss are shown in Figure \ref{fig:train_eval_loss}.

\end{appendices}

\bibliography{bibliography}

\end{document}